\documentclass[12pt,fleqn,a4paper]{article}
\usepackage{amssymb}
\usepackage{amsmath}
\begin{document}
\begin{flushright}
KEK-TH-776 \\
{\tt hep-ph/0107084 }\\
\end{flushright}
\vspace*{1.5cm}
\begin{center}
{\baselineskip 25pt
\Large{\bf 
Natural $\mu$-term generation
in supergravity scenario
}}

\vspace{1.2cm}
\def\thefootnote{\fnsymbol{footnote}}
{\large Ryuichiro Kitano$^{ab,}$\footnote
{email: {\tt ryuichiro.kitano@kek.jp}}
and Nobuchika Okada$^{a,}$\footnote
{email: {\tt okadan@camry.kek.jp}}
}
\vspace{.5cm}

{\small {\it $^a$Theory Group, KEK, Oho 1-1, Tsukuba, 
Ibaraki 305-0801, Japan \\
\vspace*{2mm}
$^b$Department of Particle and Nuclear Physics,
The Graduate University for Advanced Studies,\\
Oho 1-1, Tsukuba, Ibaraki 305-0801, Japan}}

\vspace{1.5cm}
{\bf Abstract}
\end{center}

\bigskip

We discuss a natural way to generate the $\mu$-term 
in supergravity scenario. 
Once the supergravity effects are taken into account, 
the vacuum expectation values (VEVs) of the heavy fields  
are in general shifted from the values in the supersymmetric limit. 
We note that this fact is independent of any Kahler ansatz 
and the values of the VEV shifts are 
of the order of the gravitino mass. 
As an example, an explicit model is presented, in which 
both of the $\mu$-term and the $B$-term of the electroweak scale 
are generated by the VEV shifts through the supergravity effects. 
This model is a kind of the next to minimal supersymmetric 
standard model, but there is no light standard model singlet field. 
Also, we emphasize that our discussion can be naturally applied 
to the supersymmetric grand unified theory. 


\newpage
\def\thefootnote{\arabic{footnote}}
\setcounter{footnote}{0}
\baselineskip 20pt


\section{Introduction}
Supersymmetry (SUSY) is the most attractive candidate
for the theory above TeV scale.
It ensures stability of the weak scale 
$m_W$\cite{Polchinski:1982an},
and predicts the gauge coupling unification \cite{Amaldi:1991cn}
which naturally
indicates a beautiful unification of 
the standard model gauge interactions
to the SU(5) grand unified theory (GUT) \cite{Georgi:1974sy}.
The radiative breaking of the electroweak symmetry
is also an interesting feature 
\cite{Inoue:1982ej, Alvarez-Gaume:1983gj}.

We know that SUSY is broken because
none of superpartners has not been 
observed yet.
Therefore 
we should include the SUSY breaking mechanism
in the theory.
Investigating the SUSY breaking scenario is
an important topic in particle physics
\cite{Nilles:1984ge}.
One of the successful SUSY breaking scenarios is
the supergravity scenario \cite{Chamseddine:1982jx, Hall:1983iz},
in which SUSY is broken in the hidden sector
and the SUSY breaking information communicate 
the visible sector
through the supergravity interactions.
In this scenario,
if we assume the SUSY breaking scale 
of order $10^{11}$ GeV,
soft breaking terms in the visible sector
are obtained to be of the order of weak scale
so that the electroweak scale is stabilized.

However, we have a less understanding feature 
in SUSY models,
that is so called the $\mu$-problem \cite{Kim:1984dt}.
In the minimal supersymmetric standard model (MSSM),
the superpotential have a bilinear term in the Higgs sector
such as $W \supset \mu H_1 H_2$, where
$H_1$ and $H_2$ are Higgs doublets,
and $\mu$ is a parameter of dimension one.
From the naturalness point of view,
this supersymmetric parameter $\mu$
would be of the order of the GUT ($M_{\rm GUT}$)
or Planck ($M_{\rm Pl}$) scales
or zero by some symmetry reason.
However, from a claim 
for the correct electroweak symmetry breakdown,
this $\mu$ parameter 
must be the same order
as the weak scale,
so that $\mu \sim M_{\rm GUT}$ or $M_{\rm Pl}$ is not allowed.
A vanishing $\mu$ parameter is also forbidden
because it leads to a massless chargino
being excluded by experiments \cite{Groom:2000in}.
The mysterious question is why
the ``supersymmetric'' parameter $\mu$ is 
the same order as
the weak scale whose origin is ``SUSY breaking'' 
\cite{Kim:1984dt}.

There have been many attempts towards this problem.
Giudice and Masiero considered 
a possibility of existence of
higher dimensional interaction terms
between Higgs fields and 
hidden sector fields in Kahler potential,
and these terms induce the $\mu$-term through SUSY breaking
\cite{Giudice:1988yz}.
Other models have been also considered e.g.\
so called next to minimal supersymmetric standard model
(NMSSM) \cite{Fayet:1975pd},
connecting Higgs fields with hidden sector 
in the superpotential \cite{Casas:1993mk},
and 
imposing additional symmetries \cite{Hempfling:1994ae}.

In this paper,
we consider another way to produce the $\mu$-term
in the supergravity scenario
in which 
neither
particular Kahler potential nor
particular hidden sector
is required.
We consider VEV shifts of
heavy fields through the supergravity interactions.
In the supersymmetric limit,
the VEVs for the heavy fields are 
determined by the vanishing F-term and D-term conditions.
However, when we switch on the supergravity interactions,
the potential is deformed
and VEVs for the heavy fields are shifted 
from that in the supersymmetric limit.
The values of the VEV shifts are found to be of order 
gravitino mass $m_g (\sim m_W)$ \cite{Hall:1983iz}.
We propose a scenario that
the $\mu$-term is exactly vanishing in the supersymmetric limit,
but generated by the VEV shift of heavy fields.

This paper is organized as follows:
In Section 2,
we review 
the VEV shifts in supergravity theories.
In Section 3,
we present an explicit model
of $\mu$-term generation by using
VEV shifts of heavy fields
and discuss the application to GUT theory.
In Section 4, 
we give our conclusions.

\section{VEV shifts in supergravity}
In this section,
we review VEV shifts of heavy fields
in the supergravity scenario.
The values of the VEV shifts are found
to be of the order of $m_g$ \cite{Hall:1983iz}.

The scalar potential of supergravity is given by 
\cite{Cremmer:1983en}
\begin{eqnarray}
 V = e^K
\left[
\left( \frac{\partial W}{\partial z_i}
+ \frac{\partial K}{\partial z_i} W  \right) 
g^{i \bar{j}}
\left( \frac{\partial W^*}{\partial z_j^*}
+ \frac{\partial K}{\partial z_j^*} W^*  \right) 
- 3 |W|^2
\right]
+ (\mbox{D-terms})\ ,
\label{1}
\end{eqnarray}
where we take a unit $8 \pi G_N = 1$,
$K$ and $W$ are Kahler potential and superpotential,
respectively,
$z_i$ represents scalar components of
chiral superfields, and $g^{i\bar{j}}$ is Kahler metric:
\begin{eqnarray}
 g_{i \bar{j}} = 
\frac{ \partial^2 K}{\partial z_i \partial z_j^*}
\ .
\end{eqnarray}
In supergravity scenario,
the superpotential is divided into 
visible and hidden sectors as follows:
\begin{eqnarray}
 W = W_{\rm vis} + W_{\rm hid}\ .
\end{eqnarray}
When we claim 
the SUSY is broken in the hidden sector
and the vanishing cosmological constant conditions,
$ \langle V \rangle = 0 $ and 
$ \langle W_{\rm vis} \rangle = 0 $,
the gravitino mass is given by
$ m_g = \langle e^{K/2} W_{\rm hid} \rangle 
\sim \langle W_{\rm hid} \rangle \sim m_W$.

In the visible sector,
the field VEVs in the supersymmetric limit
are determined
by the vanishing F-term condition such as
$ \partial W_{\rm vis} / \partial z_i |_{z_i = z_i^0} = 0 $.
However, note that in the potential given in eq.(\ref{1}),
the VEVs are shifted by the SUSY breaking effect.
Here, we parameterize the VEV shift
of the visible sector fields as
\begin{eqnarray}
  z_i = z_i^{0} + \delta z_i \ .
\end{eqnarray}
Assuming 
$\delta z_i \sim O(m_g) \sim \langle W_{\rm hid} \rangle$, and 
expanding the potential
with respect to 
$\langle W_{\rm hid} \rangle$ and $\delta z_i$,
the leading order terms are given by
\begin{eqnarray}
 V \simeq e^K
\left( 
\left. \frac{\partial^2 W_{\rm vis}}{\partial z_i \partial z_k} 
\right|_{z^0}
\delta z_k
+ 
\left. \frac{\partial K}{\partial z_i} \right|_{z^0}
\langle W_{\rm hid} \rangle  \right) 
g^{i \bar{j}}
\left( 
\left. \frac{\partial W^*_{\rm vis}}{\partial z_j^* \partial z_l^*}
\right|_{z^0}
\delta z_l^*
+ 
\left. \frac{\partial K}{\partial z_j^*} \right|_{z^0}
\langle W_{\rm hid}^* \rangle \right)
\ .
\end{eqnarray}
The stationarity condition 
$ \partial V / \partial (\delta z_i) = 0 $ leads to
\begin{eqnarray}
 \delta z_i = 
- \left(
\left.
\frac{\partial^2 W_{\rm vis}}{\partial z_i \partial z_j}
\right|_{z^0}
\right)^{-1}
\left.
\frac{\partial K}{\partial z_j} 
\right|_{z^0}
\ 
\langle W_{\rm hid} \rangle \ .
\label{6}
\end{eqnarray}
This is generally of the order of $m_g$.
The inversibility of 
$ {\partial^2 W_{\rm vis}}/{\partial z_i \partial z_j} |_{z^0} $
means that the fields $z_i$'s are all massive.
We can see that VEV shifts of heavy fields are
of the order of $m_g$.

\section{Models}
In this section,
we propose an explicit model 
in which the $\mu$-term is generated 
by the VEV shifts discussed in the previous section.
Although
the $\mu$-term is absent in the supersymmetric limit,
non-zero $\mu$-term emerges through the VEV shifts of heavy fields.

The visible sector superpotential is given by
\begin{eqnarray}
 W = W_{\rm MSSM} + \lambda_H S H_1 H_2 
+ \lambda_N S ( N^2 - m^2 )\ ,
\label{7}
\end{eqnarray}
where $ W_{\rm MSSM} $ is the
superpotential in MSSM except for the $\mu$-term,
$S$ and $N$ are standard model singlet chiral superfields,
$m$ is a mass parameter of order GUT or Planck scale,
and $\lambda_H$ and $\lambda_N$ are
dimensionless coupling constants.
This superpotential is general under the 
standard model gauge group and R-symmetry
where we assign R-charge 2 for $S$,
0 for $H_1$, $H_2$, and $N$, and 
1 for all other chiral superfields in MSSM.
Here we omit 
bilinear term $SN$ by the field redefinition of 
$N$.\footnote{
Although this field redefinition may induce
a linear term in the Kahler potential,
it does not change our result of $\mu$-term and $B$-term
generation.
}
In the supersymmetric limit,
the field VEVs are given by
\begin{eqnarray}
 \langle S \rangle = 0 \ ,\ 
 \langle N \rangle = m \ ,
\label{8}
\end{eqnarray}
where we assume SU(2)$_{\rm L} \times$ U(1)$_{\rm Y}$ 
unbroken vacuum i.e.\ 
$\langle H_1 \rangle = \langle H_2 \rangle = 0$.
In this stage, the Higgs doublets are massless and 
the field $S$ and $N$ are both superheavy.

Let us estimate the VEV shifts for $S$ and $N$
by substituting eq.(\ref{7}) into eq.(\ref{6}).
The second derivatives of the superpotential
(which is mass matrix for $S$ and $N$)
are given by
\begin{eqnarray}
\frac{\partial^2 W_{\rm vis}}{\partial z_i \partial z_j}
 = 
\left(
\begin{array}{cc}
 0 & 2 \lambda_N m \\
 2 \lambda_N m & 0 \\
\end{array}
\right)\ ,
\label{10}
\end{eqnarray}
where $z_1 = S$ and $z_2 = N$.
For simplicity, we assume that
the zero-th order of 
the Kahler potential in the power series of $z_i/M_{\rm Pl}$,
is of the canonical form, $ K = S^* S + N^* N $. 
The higher order terms can be neglected if 
$z_i / M_{\rm Pl} \ll 1$.\footnote{
This assumption is introduced just for simplicity.
Even if $z_i / M_{\rm Pl}$ is not small,
we can obtain suitable $\mu$-term and $B$-term proportional to $m_g$
with the coefficients which depend on Kahler potential.
}
Now
we can obtain the leading order VEV shifts from 
eqs.(\ref{6}), (\ref{8}), and (\ref{10}) as follows:
\begin{eqnarray}
 \delta S =
-\frac{1}{2 \lambda_N} \langle W_{\rm hid} \rangle \ ,\ 
 \delta N = 0\ .
\end{eqnarray}
Note that $S$ acquires the VEV of order $m_g$,
which means that the $\mu$-term of order $m_W$
is successfully generated.
More explicitly,
we can write down the low energy effective potential
by integrating out the $S$ and $N$ fields as
\begin{eqnarray}
 V_{\rm eff} \!\!\!&=&\!\!\! \lambda_H^2 | \delta S |^2 
 ( |H_1|^2 + |H_2|^2 )
- 2 \lambda_H \lambda_N 
( |\delta S|^2 + |\delta N|^2 ) H_1 H_2 + {\rm h.c.}
\nonumber \\
&&+ (\mbox{D-terms}) + (\mbox{soft terms})\ ,
\end{eqnarray}
where the last term denotes the soft terms
dependent on the Kahler potential.
In the above equation,
we can see that
the $B$-term (${\cal L} \supset -B \mu H_1 H_2$)
can be also generated, $B \mu \sim O(m_g^2)$,
which is suitable for electroweak symmetry breaking.

Note that
the low energy effective theory of our model
after integrating out 
heavy singlet fields $S$ and $N$ is
just MSSM,
and there is no light fields except for the MSSM
particle contents.
This is the crucial difference from usual NMSSM
in which there is a light singlet field
in low energy superpotential \cite{Fayet:1975pd}.
In NMSSM,
a standard model singlet field $X$ is introduced and 
couples to Higgs fields as
$W_{\rm NMSSM} \supset X H_1 H_2 + X^3$.
The minimization condition for the potential
including the soft terms leads to
$\langle X \rangle \sim m_W$
which generates effectively the $\mu$-term.
However, 
in supergravity scenario with such a light singlet field,
there is a serious problem that
the weak scale is destabilized by large tadpole operator
induced by supergravity interactions
\cite{Bagger:1993ji}
or
GUT interactions
\cite{Alvarez-Gaume:1983gj}.
The present model does not suffer from this problem
because 
the singlet fields $S$ and $N$ are
heavy enough to ignore such tadpole terms. 
In GUT models,
the $\mu$-problem is a part of 
doublet-triplet Higgs mass splitting problem.
The two Higgs doublets are embedded into
${\bf 5}$ and ${\bf \bar{5}}$ of SU(5).
The success of gauge coupling unification
requires the SU(2) doublet parts of 
${\bf 5}$ and ${\bf \bar{5}}$ are light ($\lesssim m_W$)
while SU(3) triplet parts are heavy ($\sim M_{\rm GUT}$).
In the minimal SU(5) GUT model,
this splitting is accomplished by fine-tuning of 
the parameters in the GUT breaking sector.
The superpotential of the GUT breaking sector is
given by
\begin{eqnarray}
 W = m H \bar{H} + H \Sigma \bar{H} + V(\Sigma)\ ,
\end{eqnarray}
where $H$ and $\bar{H}$ are ${\bf 5}$ and ${\bf \bar{5}}$
Higgs fields, and $\Sigma$ is an adjoint representation field.
The light doublets needs fine-tuning between 
the Higgs mass parameter $m$ and $\langle \Sigma \rangle$
with accuracy of $10^{-14}$ level.
However, if we can find a mechanism 
in which SUSY vacuum condition requires
vanishing doublet Higgs masses, namely,
a particular relation between $m$ and $\langle \Sigma \rangle$,
the $\mu$-term and $B$-term 
can be generated in the correct order
by the mechanism discussed in the previous section
\cite{Alvarez-Gaume:1983gj}.
As an example of this approach,
the model recently proposed by the present 
authors \cite{Kitano:2001ie}
is very remarkable,
in which
the doublet-triplet Higgs mass splitting is realized 
by means of the SUSY gauge dynamics in the supersymmetric limit.
We can find that both of the $\mu$-term
and $B$-term of order $m_W$ are really generated
in this model through the VEV shifts 
by the supergravity effects.\footnote{
In ref.\cite{Kitano:2001ie},
we mentioned that we cannot obtain
$\mu$-term of order $m_W$ according to dimensional analysis.
However, in supergravity scenario,
the naive analysis
is not applicable.
We can find that
the $\mu$-term of order $m_W$ is really generated in this model.
This fact makes the model more impressive.
}

\section{Conclusions}
In conclusion,
we presented explicit models
in which 
$\mu$-term is generated by
the VEV shifts of heavy fields
in the supergravity scenario.
In the supersymmetric limit,
$\mu$-term is forbidden by
the R-symmetry \cite{Kim:1994eu}
which is broken in the hidden sector.
The VEV shifts of heavy fields
are generally of order $m_g$
without any assumptions for 
a particular form of Kahler potential
or
a particular hidden sector.
This mechanism requires the presence of a heavy 
standard model singlet field
which couples to Higgs doublets.
In GUT models, it can be identified as the GUT breaking field
e.g.\ the standard model singlet component of 
the SU(5) adjoint representation field.

In GUT models,
the $\mu$-problem is connected to
the doublet-triplet Higgs mass splitting problem.
This problem 
is one of the most challenging problems
in particle physics.
The standard model requires
SUSY from the naturalness point of view
and 
GUT in order to
account for the electric charge quantization of fermions.
Although
SUSY and GUT are independently required,
SUSY surprisingly predicts the gauge coupling unification 
with the GUT normalization of the U(1)$_{\rm Y}$ charge.
Therefore
the SUSY GUT may be
a consistent and attractive theory as the high energy physics.
The remained problem is 
to naturally realize
the doublet-triplet Higgs mass splitting.
The interesting point of this problem is that
the resolution of this problem needs
its failure, namely, 
the $\mu$-term for the doublet Higgs fields must not
completely vanish,
but be
of order $m_W$.
This fact may give us a hint to solve this problem.
A natural approach for
the doublet-triplet Higgs mass splitting is that
the splitting completely successes in the supersymmetric limit,
but SUSY breaking effects disturb this success.
Our conclusion is that
such a scenario is possible
in the supergravity scenario.

\section*{Acknowledgments}
We would like to thank Yukihiro Mimura for
useful discussions.
This work was supported by the JSPS Research Fellowships
for Young Scientists (R.K.).


\newpage
\begin{thebibliography}{99}
\bibitem{Polchinski:1982an}
J.~Polchinski and L.~Susskind,
Phys.\ Rev.\ D {\bf 26}, 3661 (1982).
\bibitem{Amaldi:1991cn}
U.~Amaldi, W.~de Boer and H.~Furstenau,
Phys.\ Lett.\ B {\bf 260}, 447 (1991); \\
J.~R.~Ellis, S.~Kelley and D.~V.~Nanopoulos,
Phys.\ Lett.\ B {\bf 260}, 131 (1991); \\
P.~Langacker and M.~Luo,
Phys.\ Rev.\ D {\bf 44}, 817 (1991).
\bibitem{Georgi:1974sy}
H.~Georgi and S.~L.~Glashow,
Phys.\ Rev.\ Lett.\  {\bf 32}, 438 (1974).
\bibitem{Inoue:1982ej}
K.~Inoue, A.~Kakuto, H.~Komatsu and S.~Takeshita,
Prog.\ Theor.\ Phys.\ {\bf 67}, 1889 (1982);
Prog.\ Theor.\ Phys.\ {\bf 68}, 927 (1982);
Prog.\ Theor.\ Phys.\ {\bf 71}, 413 (1984);\\
K.~Inoue, A.~Kakuto and S.~Takeshita,
Prog.\ Theor.\ Phys.\ {\bf 71}, 348 (1984).
\bibitem{Alvarez-Gaume:1983gj}
L.~Alvarez-Gaume, J.~Polchinski and M.~B.~Wise,
Nucl.\ Phys.\ B {\bf 221}, 495 (1983).
\bibitem{Nilles:1984ge}
For review, see
H.~P.~Nilles,
Phys.\ Rept.\  {\bf 110}, 1 (1984);
M.~Dine,
hep-ph/9612389.
\bibitem{Chamseddine:1982jx}
A.~H.~Chamseddine, R.~Arnowitt and P.~Nath,
Phys.\ Rev.\ Lett.\  {\bf 49}, 970 (1982); \\
R.~Barbieri, S.~Ferrara and C.~A.~Savoy,
Phys.\ Lett.\ B {\bf 119}, 343 (1982).
\bibitem{Hall:1983iz}
L.~Hall, J.~Lykken and S.~Weinberg,
Phys.\ Rev.\ D {\bf 27} (1983) 2359.
\bibitem{Kim:1984dt}
J.~E.~Kim and H.~P.~Nilles,
Phys.\ Lett.\ B {\bf 138}, 150 (1984).
\bibitem{Groom:2000in}
D.~E.~Groom {\it et al.}  [Particle Data Group Collaboration],
Eur.\ Phys.\ J.\ C {\bf 15}, 1 (2000).
\bibitem{Giudice:1988yz}
G.~F.~Giudice and A.~Masiero,
Phys.\ Lett.\ B {\bf 206}, 480 (1988).
\bibitem{Fayet:1975pd}
P.~Fayet,
Nucl.\ Phys.\ B {\bf 90}, 104 (1975);\\
H.~P.~Nilles, M.~Srednicki and D.~Wyler,
Phys.\ Lett.\ B {\bf 120}, 346 (1983);\\
J.~M.~Frere, D.~R.~Jones and S.~Raby,
Nucl.\ Phys.\ B {\bf 222}, 11 (1983);\\
J.~P.~Derendinger and C.~A.~Savoy,
Nucl.\ Phys.\ B {\bf 237}, 307 (1984).
\bibitem{Casas:1993mk}
J.~A.~Casas and C.~Munoz,
Phys.\ Lett.\ B {\bf 306}, 288 (1993).
\bibitem{Hempfling:1994ae}
R.~Hempfling,
Phys.\ Lett.\ B {\bf 329}, 222 (1994);\\
E.~J.~Chun,
Phys.\ Lett.\ B {\bf 348}, 111 (1995); \\
Y.~Nir,
Phys.\ Lett.\ B {\bf 354}, 107 (1995).
\bibitem{Cremmer:1983en}
E.~Cremmer, S.~Ferrara, L.~Girardello and A.~Van Proeyen,
Nucl.\ Phys.\ B {\bf 212}, 413 (1983).
\bibitem{Bagger:1993ji}
J.~Bagger and E.~Poppitz,
Phys.\ Rev.\ Lett.\  {\bf 71}, 2380 (1993).
\bibitem{Kitano:2001ie}
R.~Kitano and N.~Okada,
hep-ph/0105220,
to appear in Phys.\ Rev.\ D.\
\bibitem{Kim:1994eu}
J.~E.~Kim and H.~P.~Nilles,
Mod.\ Phys.\ Lett.\ A {\bf 9}, 3575 (1994).
\end{thebibliography}
\end{document}